# Color for the perceptual organization of the pictorial plane: Victor Vasarely's legacy to Gestalt Psychology

**Birgitta Dresp-Langley[1], Adam Reeves[2]**

[1] Centre National de la Recherche Scientifique CNRS, ICube UMR 7357 CNRS -Université de Strasbourg, FRANCE; birgitta.dresp@unistra.fr

[2] Northeastern University, Psychology Department, Boston, USA; a.reeves@northeastern.edu

\*         Correspondence: birgitta.dresp@unistra.fr

**Abstract**

Victor Vasarely's (1906-1997) important legacy to the study of human perception is brought to the forefront and discussed. A large part of his impressive work conveys the appearance of striking three-dimensional shapes and structures in a large-scale pictorial plane. Current perception science explains such effects by invoking brain mechanisms for the processing of monocular (2D) depth cues. Here in this study, we illustrate and explain local effects of 2D color and contrast cues on the perceptual organization in terms of figure-ground assignments, i.e. which local surfaces are likely to be seen as "nearer" or "bigger" in the image plane. Paired configurations are embedded in a larger, structurally ambivalent pictorial context inspired by some of Vasarely's creations. The figure-ground effects the configurations produce reveal a significant correlation between perceptual solutions for "nearer" and "bigger" when other monocular depth cues are not provided. In consistency with previous findings on similar, albeit simpler visual displays, a specific color may compete with luminance contrast to resolve the planar ambiguity of a complex pattern context at a critical point in the hierarchical resolution of figure-ground uncertainty. The potential role of color temperature in this process is brought forward here. Vasarely intuitively understood and successfully exploited the subtle context effects accounted for in this paper, well before empirical investigation had set out to study and explain them in terms of information processing by the visual brain.

*"Every form is a base for color, every color is the attribute of a form"*

Victor Vasarely (1906-1997)

**Introduction**

Victor Vasarely's (1906-1997) major work was essentially inspired by Gestalt Theory (Metzger, 1930, Rubin, 1921, Wertheimer, 1923). It exploits axonometric squares or cubes with more or less curvilinear contours, to convey the appearance of a three-dimensional structure to the pictorial plane. By varying the size, luminance contrast, and/or color of the cubes, Vasarely created powerful 3D effects using minimalist variations in 2D geometry. Perception science has only recently begun to understand the brain mechanisms driving this perceptual organization of planar image data on the basis of monocular cues to 3D that were already described by Leonardo da Vinci (da Vinci, 1651), and further discussed and illustrated centuries later by the Italian Gestalt Theorist Gaetano Kanizsa (e.g. Kanizsa, 1979). Such 2D cues enable both grouping and/or segregation of specific parts of the image plane on the basis of local differences in size, luminance contrast, and/or color and thereby confer order to a multitude of simultaneously incoming visual signals (von der Heydt, 2016). Such order expresses itself in statistically significant brain representations of structural regularity with psychophysically measurable perceptual correlates (Dresp-Langley, Reeves, and Grossberg, 2017).

Two of the most powerful 2D cues to visual 3D are *relative size* and *linear perspective* (Figure 1). Of two objects in the pictorial plane, the larger one is statistically the more likely to appear nearer to the human observer than the smaller one, as shown psychophysically in systematic studies on human perception (Guibal and Dresp, 2004). This monocular depth cue of *relative size* is perceptually reinforced by additional 2D cues of *linear perspective* (Figure 1A) and/or *luminance contrast* (Figure 1B), where the object with the lower position in the plane and/or the stronger contrast will have an even higher probability to appear nearer to the human observer, especially in increasingly complex pictorial pattern contexts, where a single local cue of relative size becomes increasingly harder to detect if not supported by an additional depth cue (Dresp, Durand, and Grossberg, 2002; Guibal and Dresp, 2004). *Curvature* (Dresp-Langley, 2015) and bilateral *symmetry* (Dresp-Langley, 2016) reinforce the effects of linear perspective on 3D shape perception in the pictorial plane (Dresp-Langley, 2019). Vasarely combined these geometric 2D cues, which produce particularly salient depth effects in displays with achromatic contrast variations. Two examples, "Bianco" (1987) and "VegaIII" (1956) are shown here (Figure 1A and B, respectively).

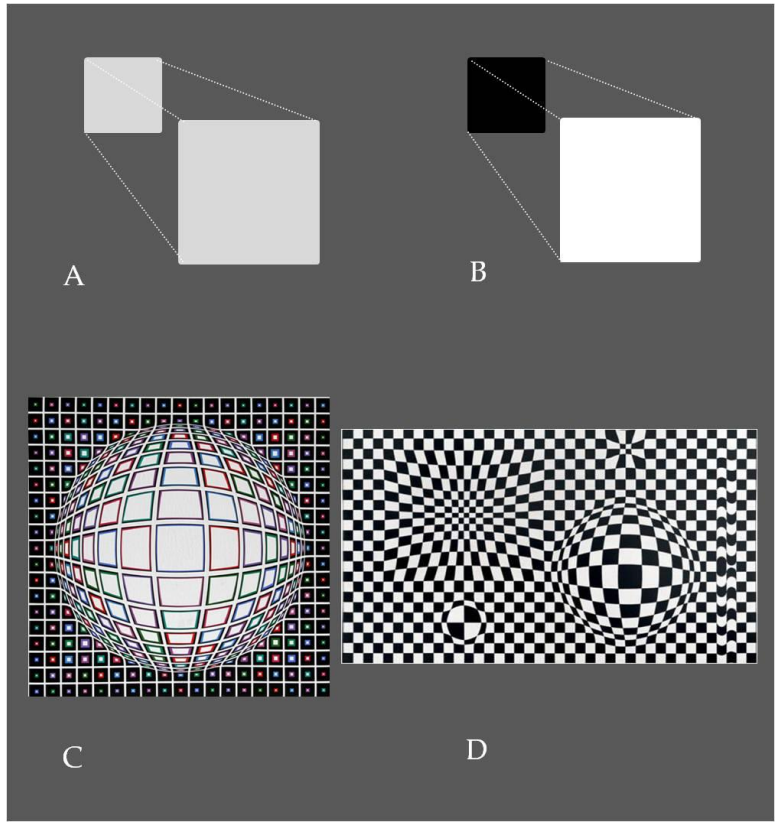

**Figure 1.** Of two objects in the pictorial plane, the larger one is the more likely to appear nearer to the human observer than the smaller one (A, B). This monocular depth cue of *relative size* is reinforced by additional 2D cues of *linear perspective* (A) and/or *luminance contrast* (B), where the object with the lower position in the plane and/or the stronger contrast will have an even higher probability to appear nearer to the human observer, especially in a more complex pictorial pattern context, where a single local cue of relative size becomes harder to detect if not supported by an additional cue. *Curvature* and bilateral *symmetry* reinforce the effects of linear perspective. Vasarely combined such 2D cues most effectively in his work, as shown here on the example of 'Bianco' (C) and 'Vega III' (D) generated from photographs taken by the first author, with permission, at the Vasarely Foundation in Aix-en-Provence, France (The Vasarely Foundation, 2020).

Other, more complex, geometric cues enabling the perception of 3D structure in pictorial displays are *contour intersection* and *partial surface occlusion*. They allow for the emergence of surfaces that convey sensations of "nearer" versus "further away" (Kanizsa, 1979, Grossberg, 1994, 1997) that can be psychophysically measured in terms of the probability of the vertical surface (Figure 2A, B, C) to appear "nearer" to a human observer (Dresp, Durand, and Grossberg, 2002, Guibal and Dresp, 2004). Line contours and adjacent groups of lines giving rise to the emergence of surfaces in the pictorial plane. In the case of spatially superposed (A, B, C) or spatially adjacent (D, E) line contours, 2D cues of local *contour intersection* (A, B) and/or *partial surface occlusion* determine, as shown in psychophysical studies, which surface is the most likely to appear nearer to the human observer. A strictly local cue of *contour intersection* (A) is perceptually reinforced by additional cues of *surface contrast* and/or *partial surface occlusion* (B, C). Spatially adjacent rows of orthogonal (D) or phase-shifted (Soriano, Spillmann, and Bach, 1996) lines (E) generate powerful cues of *intersection* and *partial occlusion* that give rise to the emergence of perceptual 3D structure, as in Vasarely's oeuvre "Riu Kiu" (1956) shown here (Figure 2F).

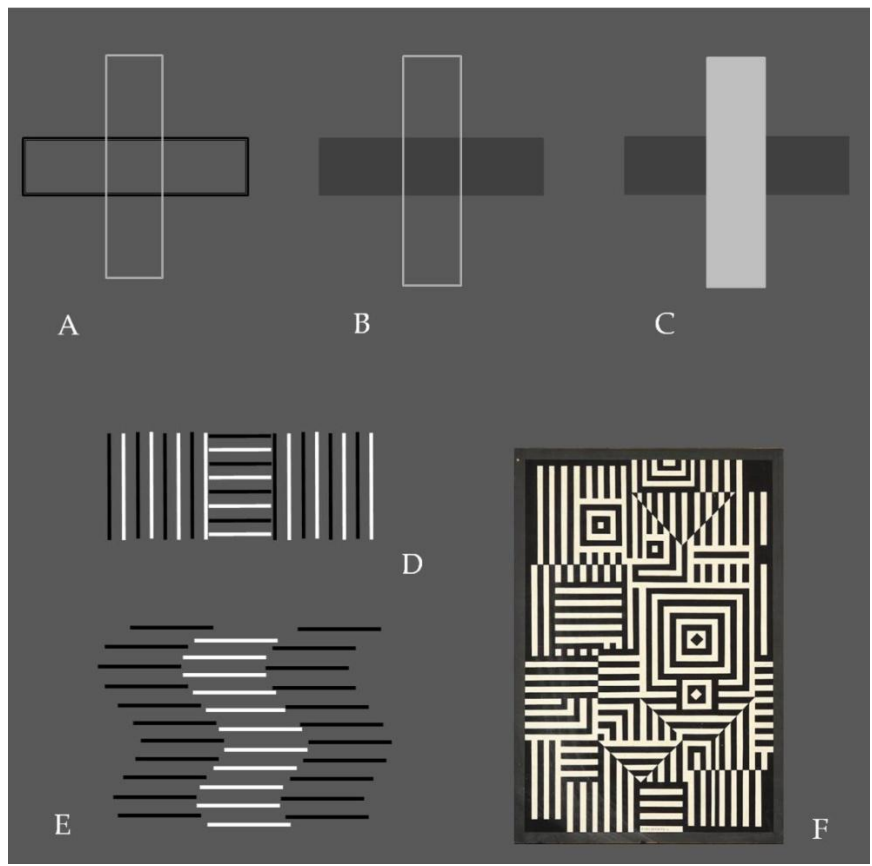

**Figure 2.** Line contours and adjacent groups of lines giving rise to the emergence of surfaces in the pictorial plane. In the case of spatially superposed (A, B, C) or spatially adjacent (D, E) line contours, 2D cues of local *contour intersection* (A, B) and/or *partial surface occlusion* determine which surface is the most likely to appear nearer to the human observer. A strictly local cue of *contour intersection* (A) is perceptually reinforced by additional cues of *surface contrast* and/or *partial surface occlusion* (B, C). Spatially adjacent rows of orthogonal (D) or phase-shifted (E) lines generate powerful cues of *intersection* and *partial occlusion* that give rise to the emergence of perceptual 3D structure. Images displayed in A-D are author generated originals. Depth cues of *intersection* and *partial occlusion* are effectively exploited in Vasarely's oeuvre "Riu Kiu" (F). The reproduction here was generated from a photograph taken by the first author, with permission, at the Vasarely Foundation in Aix-en-Provence, France (The Vasarely Foundation, 2020).

Insights from perception science have led to an understanding that any of the monocular (2D) cues to spatial structure in the pictorial plane may *cooperate* or *compete* in the genesis of "nearer", depending on the way they are combined in a complex image context. This subtle interplay, from cue cooperation to cue competition, in visual perceptual organization is accounted for by the LAMINART models, developed by the mathematician and philosopher Stephen Grossberg and his coworkers (Grossberg, 1994, 1997, 2015), on the basis of cooperative and competitive neural interactions in the visual brain, some of which were found to extend well beyond the previously assumed limits of spatial integration (for a review see Spillmann, Dresp-Langley, and Tseng, 2015). It was also found that *luminance contrast* and *color* (hue) have a particular status in these interactions. For example, a sufficiently strong *local contrast* cue competes with, and may override cues of *relative size*, *local contour intersection*, and *partial occlusion* in the perceptual genesis of "nearer" in a given image context (Guibal and Dresp, 2004). The least well studied cue to depth in the pictorial plane is color. A perceptual study by the Japanese physicist Hiroyuki Egusa (Egusa, 1983) has shown that, when no other monocular cues to depth are presented in the image plane, the color (hue) red will determine the object seen as nearer to the human observer when presented together with an achromatic object of the same contrast, shape, and size. Subsequently, it was shown that a fully saturated red may override, or win against, a green with the same luminance contrast, or a grey with a stronger luminance contrast within a restricted range of background grey levels (Guibal and Dresp, 2004). These findings suggest that, when no other 2D cues to depth are given in the plane, color, in the same way as luminance contrast (O'Shea, 1994), acquires the status of a self-sufficient monocular depth cue. This was not known to Vasarely when he created the "Arny" (*shadow*) series in the years 1966-1968. Yet, his intuition and experimentation with color combinations in the plane led

66

him to understand this important, hitherto underrated and not investigated, function of color in the perceptual organization of complex pictorial patterns. The striking depth effects he obtained are shown here (Figure 3A and B respectively) on the examples of "Arny1" (1968) and that of an untitled poster design he created in 1969 for the Thomas Gallery in Munich, Germany.

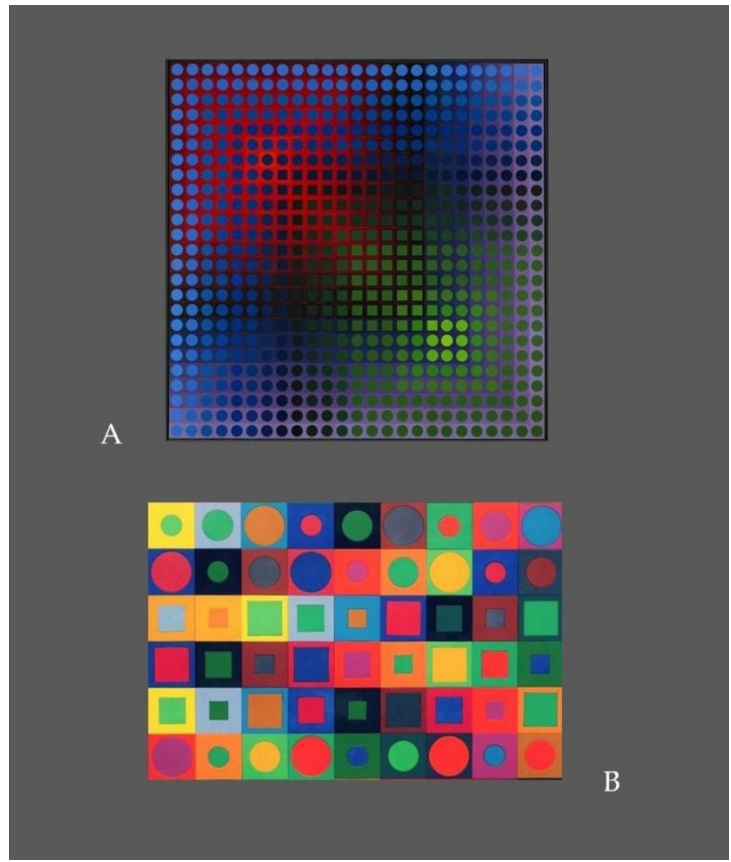

**Figure 3.** The functional role of color as a monocular cue to depth has only recently received attention from perception scientists. Vasarely's combination of variations in *color* and *luminance contrast* reveal his deep understanding of the function of color in the perceptual organization of the plane, to which he conveyed compelling effects of 3D structure that are fully appreciated when looking at the original (large scale) versions. Shown here above, in mini format for illustration: In A, the example of "Arny1" (1968), where no monocular cues to depth other than color and contrast are exploited to produce the compelling perceptual 3D structure. In B, that of an untitled poster design he created in 1969 for the Thomas Gallery in Munich, Germany (The Vasarely Foundation, 2020), where depth cues of color and contrast are supported by the variations in *relative size* of the local shapes. Reproductions shown here were generated from photographs taken by the first author, with permission, at the Vasarely Foundation in Aix-en-Provence, France (The Vasarely Foundation, 2020).

Hypotheses about how the visual brain generates these *color* and *contrast* driven 3D perceptions have been proposed in terms of non-linear antagonistic neural mechanisms, operating simultaneously to produce what is called subjective brightness assimilation and dissimilation effects (Hamada, 1985, Heinemann, 1955, Dresp *et al.*, 1991, 1996, Grossberg, 1994, 1997, 2015). Such effects are reflected by perceptions where the subjective brightness of an image background changes in a positive ("brighter") or negative ("darker") direction. Depending on the contrast polarity (negative or positive) produced by a pattern placed on the background, the induced change in perceived background brightness is termed "assimilation" when a darker pattern induces perception of a subjectively darker background and a brighter pattern induces perception of a brighter background. "Dissimilation" or subjective contrast occurs when a darker pattern induces perception of a subjectively brighter background and a brighter pattern induces perception of a subjectively darker background (Devinck *et al.*, 2006, DeWeert and Spillmann, 1995, Dresp, 1992, 1997, Dresp and Fischer, 2001, Dresp-Langley and Reeves, 2012, 2014a,b, 2018, Dresp-Langley and Grossberg, 2016, Pinna and Reeves, 2006). Results from psychophysical experiments measuring the perception of "nearer" in local shapes embedded in complex pattern contexts have allowed to point towards a functional role of these mechanisms, suggesting that they enable the coherent perceptual organization of the image plane when meaningful physical variations generating geometric depth cues are missing from the image (Dresp-Langley and Reeves, 2014a, b, 2015).

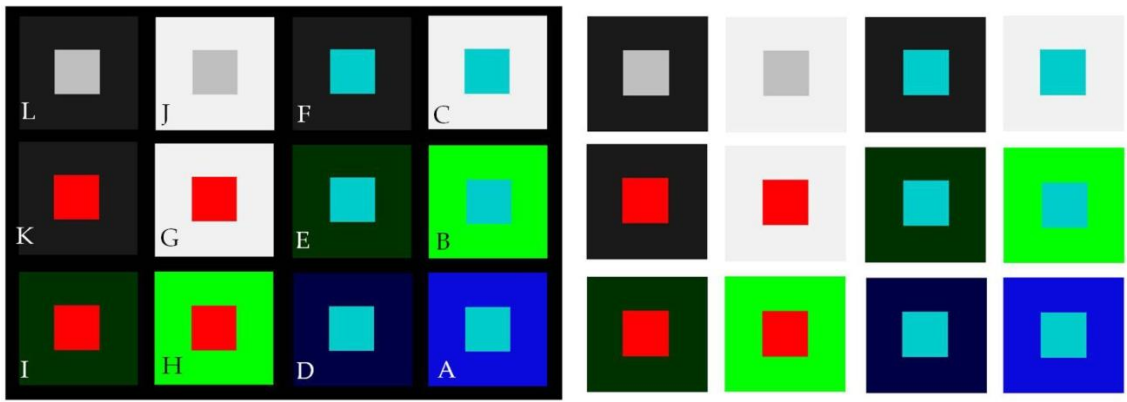

**Figure 4.** Images produced by the authors for psychophysical investigation. Local shapes were embedded in a complex pattern context, displayed against a "black wall" suggested in the pictorial plane (left), and against a "white wall" (right). Meaningful physical variations generating geometric depth cues (relative size, contour interposition, partial surface occlusion) are missing from these configurations. Local cues of *color* and *contrast* enable a coherent perceptual organization of the image plane in terms of which of two surfaces at the center of a paired configuration, numbered here from L+J to D+A, is likely to appear nearer to an observer. The corresponding color parameters (RGB, Hue °, Luminance (cd/m$^2$), Saturation%, and X, Y, Z color coordinates) are provided in the supplementary Table S3.

Here, we pursue this further by presenting results from a study where the probability of perceiving a central figural element (a red or a blue square, as shown in Figure 4) with a specific color and luminance contrast as "nearer" or as "bigger" in a pair, and as "nearest" or "biggest" in the whole, complex and variable image context. Pairs of configurations considered for assessing probabilities for "nearer" and "bigger" were L-J, K-G, I-H, F-C, E-B, and D-A, as indicated here above (Figure 1). These were paired in such a way that each pair had the same central figural element displayed on a light and on a dark local background. Combinations of pairs were displayed against a black ("black wall condition", Figure 1 left) and against a white general background ("white wall condition, Figure 1 right). The position of the pairs in the context was varied. Independent observations from independent observers were generated in counterbalanced sessions to bring to the forefront which local target color in the center of the color-surround configurations is the most likely to be seen as "nearer" and/or "bigger", which as "nearest" and/or "biggest" in the context of variations given. Presuming that contrast and color are, as predicted, self-sufficient cues to perceptual organization, the results should display intrinsically coherent probability distributions for "nearer" and "nearest" as a function of sensation magnitudes elicited by the colors in the different contexts. Whether subjective depth effects coincide with perceived differences in relative size, where one element appears subjectively "bigger" than the other in a pair, is not known but tested here. A color effect in terms of a functional advantage of the color red is expected in the light of previous findings on similar local configurations in different contexts (Egusa, 1983, Guibal and Dresp, 2004, Dresp-Langley and Reeves, 2012, 2014a, 2015).

**Results**

The cumulated responses reflecting independent observations relative to "nearer" and "bigger" of two in a pair, and/or relative to "nearest" and "biggest" in a whole image, from the ten independent observers were computed. With eight permutations of paired configurations in image displays presented in the "black wall" and "white wall conditions, we have a total of 10x8x2 independent observations.

*2.1. Descriptive analyses*

In a first analysis, the cumulated responses for "nearer" and "bigger of two in a pair" were sorted in increasing order of magnitude, without looking at which configurations produced them, and plotted as a function of the "black wall" and "white wall" presentation conditions. The results from this analysis are shown here below in Figure 5. The observed correlation between cumulated magnitudes for "nearer" and cumulated magnitudes for "bigger" in either ("black wall" and "dark wall") presentation conditions is statistically significant, as shown by the Pearson correlation coefficients (Pearson's product moment *P*) computed on the results from the two presentation conditions, with *P*=0.98 (p<.001) for the correlation between "bigger" and "nearer" in the "black wall" condition, and *P*=0.99 (p<.001) for the correlation between "bigger" and "nearer" in the "white wall" condition.

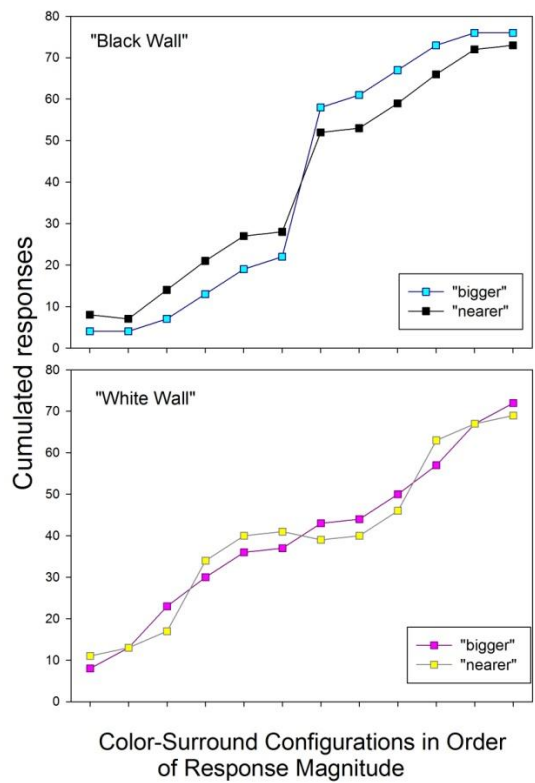

**Figure 5.** Cumulated responses for "bigger" and "nearer", produced by the twelve paired color-surround configurations (Figure 1) in independent observations, sorted here in the order of ascending response magnitudes for "bigger" associated with the corresponding magnitude for "nearer" elicited by the same display. Results from the "black wall" (top) and the "white wall" (bottom) presentation conditions are shown. A statistically significant correlation (Pearson) between the magnitudes of the responses for "nearer" and "bigger" is found.

In the next analysis, the cumulated responses for "nearer" and "bigger of two in a pair" were sorted as a function of the luminance contrast. The latter is the Weber contrast $C = (L_{center} - L_{surround})/L_{surround}$, where $L$ stands for the photometrically determined luminance. The corresponding values in $cd/m^2$ are given in the supplementary Table S3. The Weber contrasts are shown in the text here below in Table 2 of the 'Materials and Methods' section for the paired configurations which produced them. Responses were plotted as a function of the Weber contrasts for the "black wall" and "white wall" presentation conditions. The results from this analysis are shown here below in Figure 6. There is no simple (linear) function relating the response magnitudes to the physical contrast of the local color-surround configurations, as was found in some of our previous work (Dresp-Langley and Reeves, 2004a,b). However, the center-surround configurations with the weakest Weber contrasts (<10), produced the smallest magnitudes for "nearer" and "bigger" consistent with the pure contrast effects found previously with achromatic (no color) displays (Guibal and Dresp, 2004, Dresp-Langley and Reeves, 2012).

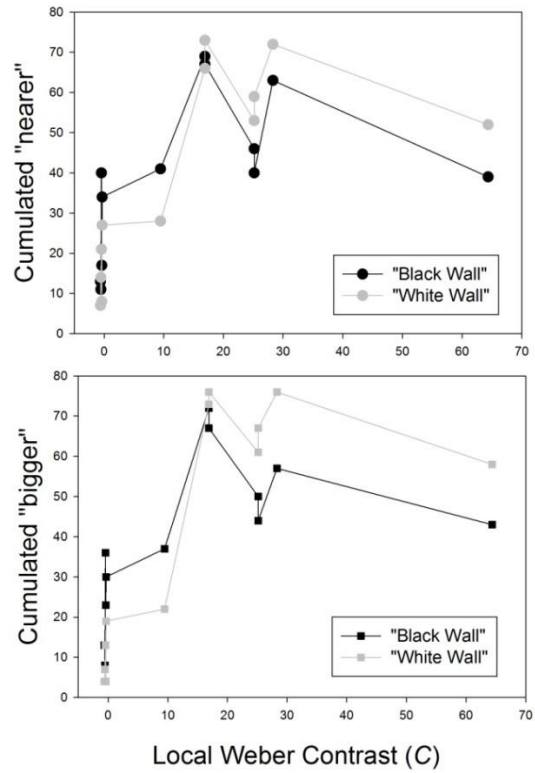

**Figure 6.** Cumulated responses for "nearer" (top) and "bigger" (bottom) sorted as a function of the luminance contrast (Weber) of the local color-surround configurations (see Table 1 here below). There is no clear function relating the response magnitudes to the physical contrast of the local center-surround configurations. However, those with the weakest Weber contrasts (<10) consistently produced the smallest magnitudes for "nearer" and "bigger".

In the next analysis, the cumulated response magnitudes produced by the local color-surround configurations (A-L) were transformed into probabilities by dividing the cumulated response magnitude by the total number of independent observations (N=80) for each local color-surround configuration. It was found that the higher cumulated response magnitudes, or probabilities, for "bigger" and "nearer" were systematically produced by a center color or tone (here red, blue and grey, as shown in Figure 4) on the darker of two surrounds in any of the six given pairs L-J to D-A (Figure 4). The corresponding probabilities of "nearer" and "bigger" were plotted as a function of the presentation condition ("black" wall and "white wall"), and are shown here below in Figure 7. It is found that, by comparison with the grey tone, the higher/highest probabilties for "nearer" and "bigger" are produced by the color red, the lower/lowest by the color blue.

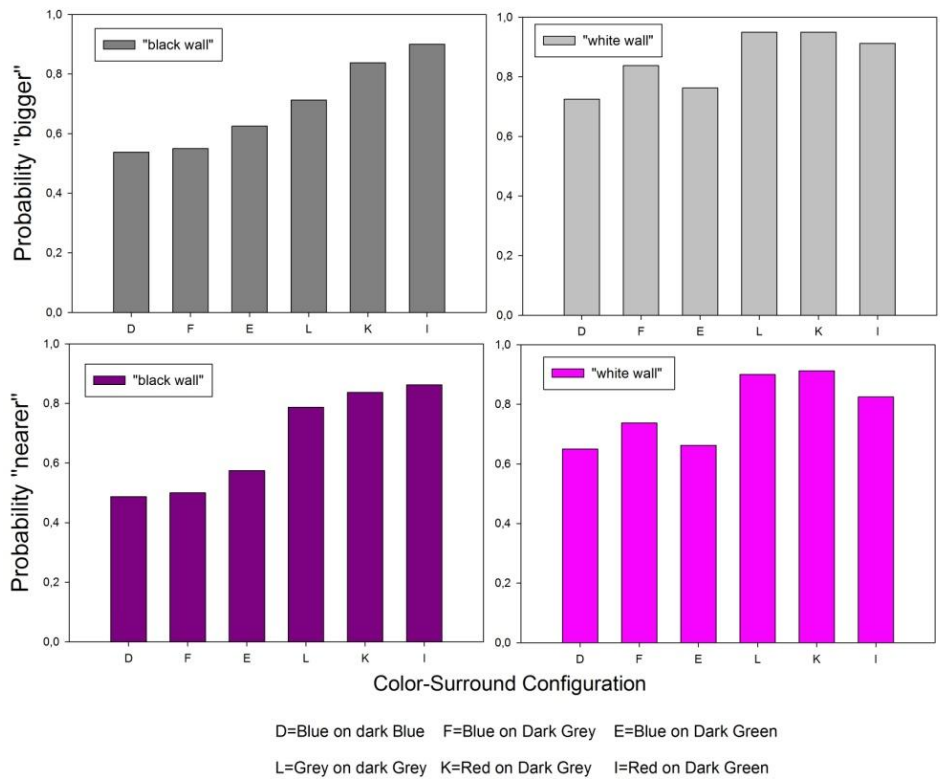

**Figure 7.** Response probabilities for "bigger" (top) and "nearer" (bottom) as a function of the local color or tone (blue, red, and grey) on the darker surround of two in a pair, and as a function of the presentation condition ("black wall" and "white wall"). The highest probabilties are produced by the color red (K and I), the lowest by the color blue (D, F and E). This probability distribution shows a clearer trend in the "black wall" condition (left) compared with the "white wall" condition (right). Response probabilities for the same color or tone on the lighter surround in a given pair are obtained by substracting each of the probabilities plotted here from 1.

The local color-surround configuration with the highest luminance contrast in the displays is blue-on-dark-blue (D), with a Weber contrast (64.38, Table 2). The local color-surround configuration with red-on-dark green (I) has a Weber contrast (16.9, Table 2) that is considerably, i.e. by more than a magnitude of three, weaker. Yet, it is found here that the probability of red-on-dark green to be seen as "nearer" or "bigger" compared with red-on-light green is significantly higher than the probability of blue-on-dark-blue compared with blue-on-light blue. This is shown by running Student's t test on the independent, cumulated response magnitude distributions produced by these two color-surround configurations in ten independent observers across independent display conditions in two independent presentation conditions ("black wall" and "white wall"). The results of this analysis are shown here below in Figure 8.

In the final descriptive analysis, the probabilities of a given color/tone in a given surround configuration to be seen as "nearest" and/or as "biggest" of all in a given image display were sorted in the order of probability magnitudes, and plotted as a fucntion of the color-surround configuration the color/tone was embedded in, and as a function of the presentation condition ("black wall" or "white wall"). The results from this analysis are shown here below in Figure 9. It is shown that the color red on the darker rey and/or green surrounds clearly stands out as the "winner", yielding the highest probabilities for relative subjective depth ("nearer") and relative subjective size ("bigger").

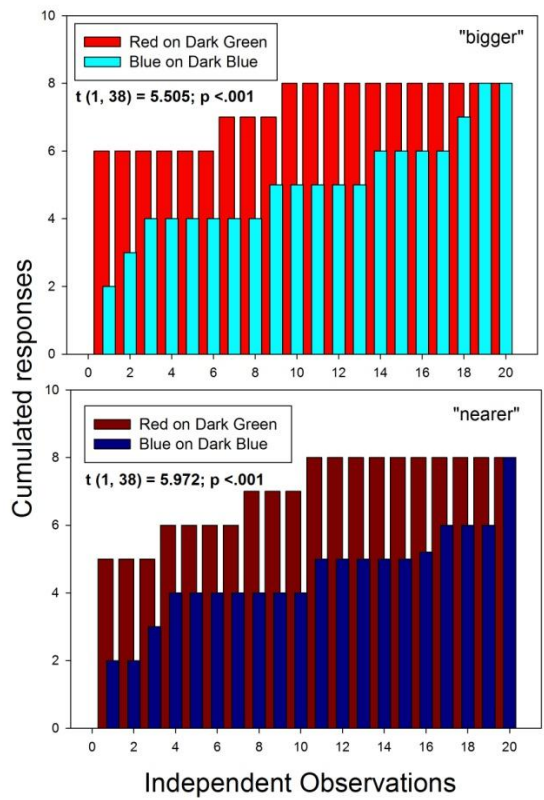

**Figure 8.** Cumulated response magnitudes for "bigger" (top) and "nearer" (bottom) sorted in the order of response magnitudes for *red-on-dark-green* (I, as in Figure 4) and *blue-on-dark-blue* (D, as in Figure 4). Despite the fact that D has a local Weber contrast (Table 1) more than three times higher than I, the latter produces significantly higher probabilities for "bigger" and "nearer".

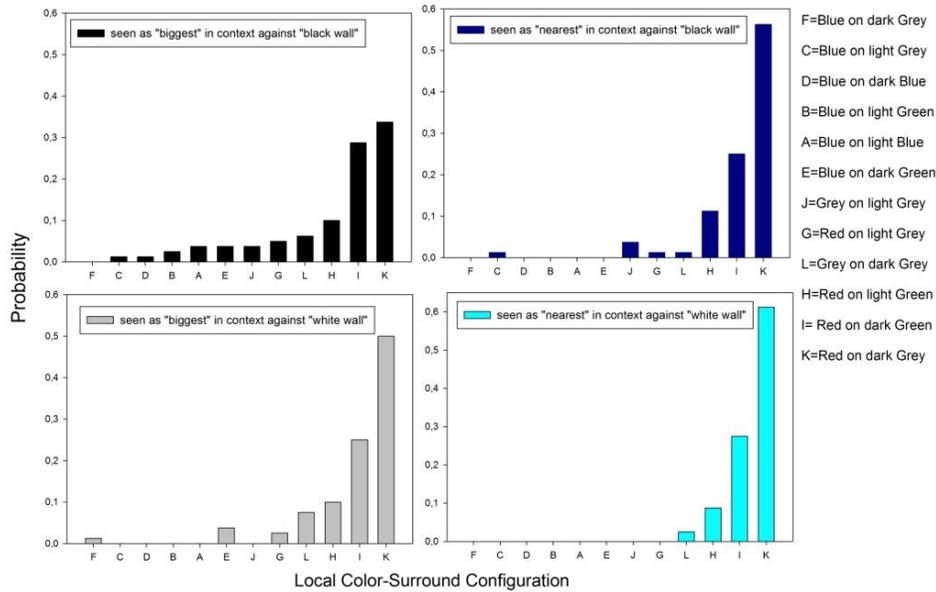

**Figure 9.** Response probabilities for a local color in a given surround to be seen as "biggest" (left) and/or "nearest" (right), plotted as a function of the presentation in the "black wall" (top) and "white wall" (bottom) conditions. The color red on the darker surrounds clearly stands out as the "winner", yielding the highest probabilities for relative subjective depth ("nearer") and relative subjective size ("bigger").

*2.2. Analysis of variance*

Further statistical analyses were performed to bring the most significant effects here to the forefront. ANalysis Of VAriance (ANOVA) was run on the cumulated response distributions for the configurations F, E, I and K (blue-dark-grey/green and red-on-dark-grey/green respectively). With ten independent observers, two independent target colors (blue, red) on two independent surrounds (dark grey, dark green), and two independent presentation ("black wall", "white wall") conditions, we have a 10x2x2x2 analysis design for distributions relative to subjective depth ("nearer") and subjective size ("bigger") effects. The results of the Two-Way ANOVA reveal statistically significant differences between the target colors red and blue for response magnitudes relative to "nearer" and "bigger" ($F(1, 79) = 33.92$, $p<.001$ and $F(1, 79) = 29.37$; $p<.001$ respectively). They also reveal significant differences between the "black wall" and "white wall" conditions on response magnitudes for "nearer" and "bigger" ($F(1, 79) = 4.35$; $p<.05$ and $F(1, 79) = 13.05$; $p<.001$ respectively). Significant interactions between variables considered in this analysis were not found. The detailed ANOVA results in terms of effect sizes, least square means and their standard errors, and post-hoc power tests are provided in the supplementary Tables S1 and S2.

**Discussion**

The effect of the luminosity of the general background on local percepts of "nearer" and "bigger" found here is consistent with other findings on display complexity, sometimes also called articulation. Articulation refers to the proximity of a perceptual target, or reference surface, with higher/lower coplanar luminance, and/or to the spatial distribution of fields of illumination in complex displays. Studies testing predictions of coplanar spatial organization have shown, for example, that articulation can substantially increase local depth effects (Radonjić and Gilchrist, 2013), as is found here. Others have shown that the variability of color judgments decreases when 3D target cubes are embedded in a constant illumination background (Allred and Olkkonen, 2013). This suggests that in the real-world, or in photorealistic 3D scenes, adding uniform backgrounds strategically can improve the stability of color and related percepts.

The finding that red surfaces yield significantly higher probabilities for "nearer" and "bigger" compared with the blue surfaces here is consistent with previous findings by the authors in configurations with several surface patches of the same color against uniform gray backgrounds (e.g. Dresp-Langley and Reeves, 2015). The previous study had shown that this functional advantage of the color red disappears when the red is sufficiently desaturated. Here in this study, all colors displayed were fully saturated, and difference in luminance contrast does not account for the color effect here either (see the parameters in Table S3 in the supplementary materials). The most likely explanation of the statistically higher probabilities for "nearer" and "bigger" produced by the red surfaces here is one in terms of color temperature. An early study from the 1950ies had shown that cold color temperatures in the range between 5400 and 6300 Kelvin require different physical adjustments to yield subjective equality compared with warmer colors in the range below 5000 Kelvin (Harrington, 1954). This suggests that cold and warm colors are perceived as qualitatively distinct, with quantifiable psychophysical correlates. It was also found that the difference in perceptual effects elicited by warm and cold colors decreases with the age of the observers. Such effects of ageing on artists' use of color is often reflected by a tendency, as they grow older, to make more extensive use of vibrant warm colors, as brought forward by Werner (1998) on the example of Monet. The effect of color temperature on our perception may be ecologically driven and affect, directly or indirectly, the subjective quality of a color. Ecologically relevance, or color quality ('attractive', 'repelling' etc.) could determine a color's probability to be seen as "nearer" in a context where decision for action in response to what is seen is mandatory. Magnitudes of foreground effects for red and blue on a dark gray from this study in comparison with results from a previous one (Dresp-Langley and Reeves, 2015) with the same colors, but different displays and task constraints, are shown in Table 1 below. In the previous study, we tested for effects of true colors on gray to validate implicit assumptions relative Chevreul's (1939) laws of color and contrast. Observers had to compare complex multiple surface patterns of the same color displayed across light and dark gray backgrounds presented sequentially in grouped and ungrouped pairs. Average probabilities for a foreground effect ("nearer") associated with the colors blue and red on dark gray backgrounds are given as a function of RGB and other color parameters including color temperature in Kelvin. Selected reds and blues on dark gray had similar saturation levels and Weber contrasts. The difference in Kelvin between the red and blue is represented by values in the extreme lower and upper regions of the Kelvin scale (°), where higher values (°) stand for colder colors, and lower values (°) for warmer colors.

Table 1: Color parameters of red and blue across two different studies and associated average probabilities of foreground effects

| *Color* | **R** | **G** | **B** | *Hue* | *L(cd/m$^2$)* | *S(%)* | **Kelvin(•)** | *Weber C* | *P "near"* |
|---|---|---|---|---|---|---|---|---|---|
| *present study single squares* | | | | | | | | | |
| RED | **255** | **0** | **0** | 0 | 35.8 | 100 | <1000 | 25.1 | 0.9 |
| BLUE | **0** | **205** | **205** | 180 | 52.3 | 100 | >6000 | 16.9 | 0.6 |
| *2015 study* | | | | | | | | | |

*multiple squares*

| | | | | | | | | | |
|---|---|---|---|---|---|---|---|---|---|
| RED | **255** | **0** | **0** | 0 | 35.8 | 100 | <1000 | 25.1 | 1 |
| BLUE | **0** | **0** | **255** | 240 | 1.5 | 100 | >6000 | 8.2 | 0.5 |

Vasarély himself exploited such effects of color temperature, placing warm against cold strategically in the plane to generate local variations in relative depth where the warmer colors appear nearer to the observer than colder tones. Photographers also exploit color temperature in this way to produce similarly stunning perceptual effects, as shown here in Figure 10.

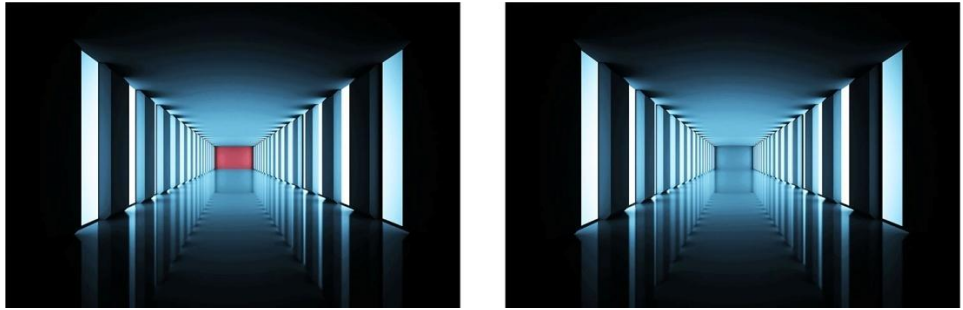

**Figure 10.** Warm red *versus* cold blue in photographically captured 3D scene effects. All other cues to 3D (linear perspective, contrast, other scene parameters) being equal in the two photographs here, the warmer red wall on the left appears nearer to us compared with the colder blue wall on the right (images shown here are openly available online; see the reference to Palad, 2013).

As pointed out by Grossberg and Zajac (2017), the human brain starts by responding to all image properties imperfectly first, and then uses hierarchical resolution of uncertainty whenever the sensitivity of local detectors is insufficient to make sense of intrinsically ambivalent physical image data. Specific multi-level interactions between detectors accomplish such hierarchical resolution of uncertainty in response to physical input, thereby generating the visual percepts that we consciously see. The general conclusion here is that the brain does not eliminate uncertainty too soon, but takes advantage of uncertainty until processing stages are reached at which uncertainty can profitably and drastically be reduced or eliminated. In the case of figure-ground perception, when all other cues to visual depth in two planar configurations are equal, the difference in temperature of a local color may become a strategic cue to pictorial depth, as successfully captured and quantified in this work here.

**Conclusions**

Vasarely's oeuvre (1906-1997) set the standard for contemporary Op Art and graphical design, and is a source of inspiration for current virtual reality applications aimed at conveying realistic three-dimensional qualities through non-dimensonal representation projection) of colour and form in affine space. The configurations from this study, embedded in larger, structurally ambivalent pictorial contexts, were inspired by some of Vasarely's creations (Figure 11), which are choice examples of powerful effects of subjective variation in relative size and depth, governing the perceptual emergence of *Gestalten* by visual organization of ambivalent image material into coherent representations of figure and ground. The statistically significant correlation between perceptual solutions for "nearer" and "bigger" found here in this study when no other monocular depth cues are given in the planar image context is the quantitative correlate of such visual organization. A warm red at the center of configurations yields statistically higher probabilities for "nearer" and "bigger" in comparison with a cold blue, suggesting that colour temperature has an important functional role at a critical point in the hierarchical perceptual resolution of figure-ground uncertainty in complex 2D patterns. Color perception is a result of evolution (Yokoyama, Xing, Liu, Faggionato, Altun and Starmer, 2014), not exclusive to humans, and subject to differences across cultures (Persaud, Hemmer, Kidd and Piantadosi, 2017). Also, species other than the human exploit color information to survival specific purposes (Dresp and Langley, 2006). In man, as the ultimate expression of this evolution, color perception is also linked to preferences, moods and emotions (Valdez and Mehrabian, 1994). In context dependent shape perception, color may compete with luminance contrast in resolving the planar ambiguity of a complex image context. As suggested previously, there is no clear function relating subjective response magnitudes for relative depth, or size, to the physical contrast of local color-surround configurations in complex displays. Yet, configurations with the weakest luminance contrasts are consistently the most likely to produce weaker depth effects. Pure contrast effects were found previously mostly with achromatic (no color) displays. The local color-surround configuration with the highest luminance contrast in the displays here, the cold blue-on-dark-blue, loses out against the warmer red-on-dark green, although the latter has a three times lesser Weber contrast.

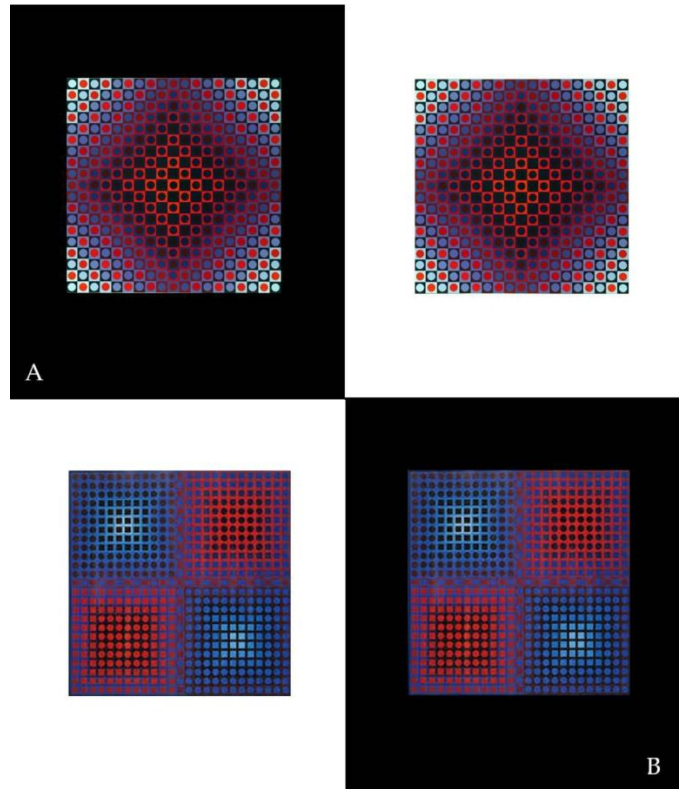

**Figure 11.** Warm red versus cold blue, in complex contrast variations against black and white backgrounds. Vasarely's "Permutations" series '(two untitled examples are shown here) illustrate the artists astonishing capacity to exploit color for perceptual figure-ground organization of planar shapes. Reproductions shown here were generated from photographs taken by the first author, with permission, at the Vasarely Foundation in Aix-en-Provence, France (The Vasarely Foundation, 2020).

Finally, any local effect of relative depth and/or size in a complex pictorial context depends on the luminosity and complexity of the display backgrounds it is presented on. This is quantified here in terms of the significant difference between the "black wall" and the "white wall" presentation conditions. Vasarely intuitively understood, and successfully exploited, all of the perceptual context effects described here.

**Materials and Methods**

General aspects of the measurement approach and underlying rationale are well documented in previously published work (Guibal and Dresp, 2004, Dresp-Langley and Reeves, 2012, 2014, 2019). Here, we specify the materials and methods from which new analyses shown and discussed in the sections here above were drawn. The images (Figure 4) were computer generated and displayed on a high resolution color monitor (EIZO COLOR EDGE CG 275W, 2560x1440 pixel resolution) connected to a DELL computer equipped with a high performance graphics card (NVIDIA). Color and luminance calibration of the RGB channels of the monitor was performed using the appropriate Color Navigator self-calibration software, which was delivered with the screen and runs under Windows 7. RGB increments were controlled using ADOBE RGB in Photoshop 7. All luminance levels were cross-checked with an external photometer (OPTICAL, Cambridge Research Systems). The physical parameters (RGB, Hue °, Luminance ($cd/m^2$), Saturation%, and X, Y, Z color coordinates) of the colors displayed here are provided in the supplementary Table S3.The size of each of the square surfaces in the center of each of the twelve local configurations in the image panels was 160x160 pixels and the size of each of the square surrounds was 400x400 pixels. The twelve local configurations were equally spaced, with 50 pixels between their surrounds, along the horizontal and vertical dimensions. The paired configurations in context are shown here above in Figure 4, displayed on black ("black wall" condition) and white ("white wall" condition) general backgrounds. The images were displayed centrally on a black or white general background of the 2560x1440 pixel screen. The size of a single pixel on the screen is 0.023 cm. Grey, red, and blue-green center squares on lighter and darker immediate surrounds were presented in pairs. Their position (left, right) in a pair as well as the position of each given pair in the global context was varied between presentations, sessions, and observers. Perceptually relevant Weber contrasts for each hue combination in the twelve paired configurations are given here below in Table 2.

**Table 2.** Center-surround hue combinations and their luminance contrast (Weber) for the twelve paired local configurations in order of decreasing contrast magnitude

| *Hue Combination* | *Weber Contrast* |
|---|---|
| Blue on Dark Blue "D" | 64.38 |
| Grey on Dark Grey "L" | 28.30 |
| Blue on Dark Grey "F" | 25.15 |
| Blue on Dark Green "E" | 25.15 |
| Red on Dark Grey "K" | 16.90 |
| Red on Dark Green "I" | 16.90 |
| Blue on Light Blue "A" | 9.46 |
| Blue on Light Green "B" | -0.33 |
| Grey on Light Grey "J" | -0.38 |
| Blue on Light Grey "C" | -0.45 |
| Red on Light Green "H" | -0.54 |
| Red on Light Grey "G" | -0.62 |

Presentation on white ("white wall" condition) and black ("black wall" condition) general backgrounds was also counterbalanced between sessions and observers. The results shown here above originate from perceptual judgments of ten independent human adult volunteer observers with normal or corrected-to-normal vision from a culturally homogenous study population of young Europeans. A sample of ten is usually sufficient to yield statistically significant effects in strong perceptual phenomena, as found in all our previous studies on this topic. In such a case, it is neither necessary nor useful to include more subjects. To do so would be mandatory when looking for cultural differences in perception, which were not investigated here. The image variations were presented to observers individually and in separated sessions, with breaks between sessions. Each individual observer was tested for normal color vision using the printed version of the Ishihara Plates (Ishihara, 1917). After adaptation to the semi-dark room for ten minutes, the observer was comfortably seated in a semi-dark room in front of the EIZO monitor at a viewing distance of about one meter. Each individual observer received standard instructions for each perceptual judgment relative to "nearer" or "nearest" and "bigger" or "biggest". Image presentation time was not limited. After a perceptual decision was made by typing the letter of the configuration chosen on the keyboard, the next image from a sequence appeared on the screen one second later. The order of variable image contexts within and between sessions was varied in such a way that no single observer saw exactly the same context more than once in a session.

**Acknowledgments:** Permission of the Vasarely Foundation in Aix-en-Provence to take pictures *in situ* is gratefully acknowledged. Examples of Vasarely's artwork shown in this article were all generated on the basis of photographs taken by B.D.L.

**References**

Allred Sarah, Olkkonen Maria. 2013.The effect of background and illumination on color identification of real, 3D objects. *Frontiers in Psychology*, 11(4), 821.

Chevreul, Michel Eugène. 1839 . De la Loi du Contraste Simultané des Couleurs. Paris: Pitois-Levrault [trans. by F. Birren (1967) as The Principles of Harmony and Contrast of Colors and their Applications to the Arts.] New York: Van Nostrand Reinhold Company.

Da Vinci, Leonardo. 1651. *Trattato della Pittura di Leonardo da Vinci*. Scritta da Raffaelle du Fresne, Langlois, Paris.

Devinck, Frédéric, Spillmann, Lothar, and Werner, Jack. 2006. Spatial profile of contours inducing long-range color assimilation. *Visual Neuroscience*, 23, 573–577.

De Weert, Charles and Spillmann, Lothar. 1995. Assimilation: Asymmetry between brightness and darkness. *Vision Research*, 35, 1413–1419.

Dresp, Birgitta, Lorenceau, Jean and Bonnet, Claude. 1990. Apparent brightness enhancement in the Kanizsa square with and without illusory contour formation. *Perception*, 19, 483-489.

Dresp, Birgitta. 1992. Local brightness mechanisms sketch out surfaces but do not fill them in: evidence in the Kanizsa Square. *Perception & Psychophysics*, 52, 562-570.

Dresp, Birgitta. 1997. On illusory contours and their functional significance, *Current Psychology of Cognition*, vol. 16(4), pp. 489–518.

Dresp, Birgitta and Grossberg, Stephen. 1999. Spatial facilitation by colour and luminance edges: boundary, surface, and attention factors. *Vision Research*, 39, 3431-3443.

Dresp, Birgitta and Fischer, Stéphane. 2001. Asymmetrical contrast effects induced by luminance and colour configurations", *Perception & Psychophysics*, 63, 1262-1270.

Dresp, Birgitta and Langley, Osmund Keith. 2005. Long-range spatial integration across contrast signs: A probabilistic mechanism? *Vision Research,* 45, 275–284.

Dresp, Birgitta and Langley, Keith. 2006. Fine structural dependence of ultraviolet reflections in the King Penguin (*Aptenodytes Patagonicus*) beak horn. *The Anatomical Record, A,* 288, 213−222.

Dresp, Birgitta, Salvano-Pardieu, Véronique and Bonnet, Claude. 1996. Illusory form from inducers with opposite contrast polarity: Evidence for multi-stage integration. *Perception & Psychophysics*, 58, 111-124.

Dresp, Birgitta, Durand, Séverine and Grossberg, Stephen. 2002. Depth perception from pairs of overlapping cues in pictorial displays, *Spatial Vision*, 15, 255–276.

Dresp-Langley, Birgitta. 2015. 2D Geometry Predicts Perceived Visual Curvature in Context-Free Viewing. *Computational Intelligence and Neuroscience*, article 708759.

Dresp-Langley, Birgitta. 2015. Principles of perceptual grouping: implications for image-guided surgery", *Frontiers in Psychology*, 6, article 1565.

Dresp-Langley, Birgitta. 2016. Affine Geometry, Visual Sensation, and Preference for Symmetry of Things in a Thing. *Symmetry*, 8, 127.

Dresp-Langley, Birgitta. 2019. Bilateral Symmetry Strengthens the Perceptual Salience of Figure against Ground. *Symmetry*, *11*, 225.

Dresp-Langley, Birgitta and Reeves, Adam. 2012. Simultaneous brightness and apparent depth from true colors on grey: Chevreul revisited, *Seeing and Perceiving*, 25, 597-618.

Dresp-Langley, Birgitta and Reeves, Adam. 2014a. Color and figure-ground: From signals to qualia", In S. Magnussen, M. Greenlee, J. Werner, A. Geremek (Eds.): *Perception beyond Gestalt: Progress in Vision*

*Research*. Psychology Press, Abingdon (UK), pp. 159-71.
Dresp-Langley, Birgitta and Reeves, Adam. 2014b. Effects of saturation and contrast polarity on the figure-ground organization of color on gray, *Frontiers in Psychology*, 5, article1136.
Dresp-Langley, Birgitta and Grossberg, Stephen. 2016. Neural Computation of Surface Border Ownership and Relative Surface Depth from Ambiguous Contrast Inputs, *Frontiers in Psychology*, 7, article1102.
Dresp-Langley, Birgitta, Reeves, Adam and Grossberg, Stephen. 2017. Editorial: Perceptual Grouping—The State of The Art. *Frontiers in Psychology*, 8, article 67.
Dresp-Langley, Birgitta *and* Reeves, Adam. 2018. Colour for behavioural success, *i-Perception*, 9(2), 1–23.
Egusa, Hiroyuki. 1983. Effects of brightness, hue, and saturation on the perceived depth between adjacent regions in the visual field. *Perception*, vol. 12, pp. 167-175, 1983.
Erlikhman, Gennady, and Kellman, Philip. 2016. From Flashes to Edges to Objects: Recovery of Local Edge Fragments Initiates Spatiotemporal Boundary Formation. *Frontiers in Psychol*ogy, 7, article 910.
Grossberg, Stephen. 1994. 3D vision and figure-ground separation by visual cortex, *Perception and Psychophysics*, 55, 48-120.
Grossberg, Stephen. 1997. Cortical dynamics of 3-D figure-ground perception of 2-D pictures. *Psychological Review*, *104,* 618-658.
Grossberg, Stephen. 2015. Cortical dynamics of figure-ground separation in response to 2D pictures and 3D scenes: How V2 combines border ownership, stereoscopic cues, and Gestalt grouping rules, *Frontiers in Psychology*, 6, article 02054.
Grossberg, Stephen and Zajac, Lauren, 2017. How humans consciously see paintings and paintings illuminate how humans see. *Art & Perception*, 5, 1-95.
Guibal, Christophe and Dresp, Birgitta. 2004. Interaction of color and geometric cues in depth perception: When does "red" mean "near"? *Psychological Research*, 10, 167-178.
Hamada, Jiro. 1985. Asymmetric lightness cancellation in Craik-O'Brien patterns of negative and positive contrast. *Biological Cybernetics*, 52, 117–122.
Harrington. 1954. Effect of color temperature on apparent brightness. *Journal of the Optical Society of America*, 44(2), 113-6.
Heinemann, Eric G. 1955. Simultaneous brightness induction as a function of inducing and test-field luminance. *Journal of Experimental Psychology*, 50, 89–96.
Ishihara, Shinobu. 1917. Tests for color-blindness (Handaya, Tokyo, Hongo Harukicho).
Kanizsa, Gaetano.1979. *Organization in vision: Essays on Gestalt perception*; Praeger: New York.
Metzger, Werner. 1930. *Gesetze des Sehens*, English trans. L. Spillmann (2009) *Laws of Seeing* Cambridge, MA: MIT Press.
O'Shea, Robert et al. 1994. Contrast as a depth cue. *Vision Research*, 34, 1595–1604.
Palad, Vinci. 2013. A landscape photographer's guide to color theory. Accessible online at: https://pixelsandwanderlust.com/landscape-photographers-guide-to-color-theory/ (last accessed on April 14, 2020).
Persaud, Kimele, Hemmer, Pernille, Kidd, Celeste and Piantadosi, Steven. Seeing Colors: Cultural and Environmental Influences on Episodic Memory. *Iperception*, 8(6), article 2041669517750161.
Pinna, Baingio and Reeves, Adam. 2006. Lighting, backlighting, and the laws of figurality in the watercolor illusion. *Spatial Vision*, 19, 341–373.
Radonjić Ana, Gilchrist Alan. 2013. Depth effect on lightness revisited: The role of articulation, proximity and fields of illumination. *I-perception*, 14, 4(6), 437-55.
Rubin, Edgar. 1921. *Visuell Wahrgenommene Figuren: Studien in psychologischer Analyse*. Kopenhagen: Gyldendalske.
Silvestri, Chiara, Motro, René, Maurin, Bernard and Dresp-Langley, Birgitta. 2010. Visual spatial learning of complex object morphologies through the interaction with virtual and real-world data. *Design Studies*, 31, 363–381.
Spillmann, Lothar and Dresp, Birgitta. 1995. Phenomena of illusory form: Can we bridge the gap between levels of explanation? *Perception*, 24, 1333-1364.
Spillmann, Lothar, Dresp-Langley, Birgitta and Tseng, Chia-Huei. 2015. Beyond the classic receptive field: The effect of contextual stimuli, *Journal of Vision*, 15, article7.

Tzvetanov, Tzvetomir and Dresp, Birgitta. 2002. Short- and long-range effects in line contrast detection. *Vision Research,* 42, 2493-2498.

Valdez, Patricia, Mehrabian, Albert. 1994. Effects of color on emotions. *Journal of Experimental Psychology General*, 123(4),394-409.

Von der Heydt, Rüdiger. 2015. Figure–ground and the emergence of proto-objects in the visual cortex, *Frontiers in Psychology,* 6, 1695, 2015.

Werner, John S. 1998. Ageing through the eyes of Monet. In: *Color Vision: Perspectives from Different Disciplines.* Walter de Gruyter & Co, Berlin-NewYork, pp.3-41.

Wertheimer, Max. 1923. *Perceived Motion and Figural Organization*, English trans. L. Spillmann *et al.* (2012) Cambridge, MA: MIT Press.

The Vasarely Foundation. 2020. Aix-en-Provence, France https://www.fondationvasarely.org/?lang=en

Yokoyama, Shozo, Xing, Jinyi, Liu, Yang, Faggionato, Davide, Altun Ahmet and Starmer William T. 2014. Epistatic adaptive evolution of human color vision. *PLoS Genetics*,10(12):e1004884.